\documentclass[12pt,preprint]{aastex}

\usepackage{rotating}
\usepackage{hyperref}
\usepackage[all]{hypcap}
\hypersetup{
    bookmarks=true,         
    unicode=false,          
    pdftoolbar=true,        
    pdfmenubar=true,        
    pdffitwindow=false,     
    pdfstartview={FitH},    
    pdftitle={My title},    
    pdfauthor={Author},     
    pdfsubject={Subject},   
    pdfcreator={Creator},   
    pdfproducer={Producer}, 
    pdfkeywords={keyword1} {key2} {key3}, 
    pdfnewwindow=true,      
    colorlinks=true,        
    linkcolor=blue,         
    citecolor=blue,         
    filecolor=magenta,      
    urlcolor=cyan           
}

\slugcomment{}

\shorttitle{Filament Channel Interaction}
\shortauthors{Navin Chandra Joshi et al.}

\begin{document}


\title{Interaction of Two Filament Channels of Different Chiralities}


\author{{Navin Chandra Joshi\altaffilmark{1}, Boris Filippov\altaffilmark{2}, Brigitte Schmieder\altaffilmark{3}, Tetsuya Magara\altaffilmark{1,4}, Young-Jae Moon\altaffilmark{1,4}}, Wahab Uddin\altaffilmark{5}}
    \affil{$^1$ School of Space Research, Kyung Hee University, Yongin, Gyeonggi-Do, 446-701, Korea; navin@khu.ac.kr, njoshi98@gmail.com}
    \affil{$^2$ Pushkov Institute of Terrestrial Magnetism, Ionosphere and Radio Wave Propagation of the Russian Academy of Sciences (IZMIRAN), Troitsk, Moscow 142190, Russia}
    \affil{$^3$ LESIA, Observatoire de Paris, PSL Research University, CNRS, Sorbonne Universit\'es, UPMC Univ. Paris 06, Univ. Paris Diderot, Sorbonne Paris Cit\'e, 5 place Jules Janssen, F-92195 Meudon, France}
    \affil{$^4$ Department of Astronomy and Space Science, Kyung Hee University, Yongin, Gyeonggi-Do, 446-701, Korea}
    \affil{$^5$ Aryabhatta Research Institute of Observational Sciences (ARIES), Manora Peak, Nainital 263 002, Uttarakhand, India}

\begin{abstract}
We present observations of interactions between the two filament 
channels of different chiralities and associated dynamics that 
occurred during 2014 April 18 -- 20. While two flux ropes of 
different helicity with parallel axial magnetic fields can only undergo a bounce interaction when they 
are brought together, the observations at the first glance show that the 
heated plasma is moving from one filament channel to the other. 
The \textit{SDO}/AIA 171 \AA\ observations and the PFSS magnetic 
field extrapolation reveal the presence of fan-spine 
magnetic configuration over the filament channels with a null
point located above them. Three different events of filament
activations, partial eruptions, and associated filament channel
interactions have been observed. The activation initiated in one
filament channel seems to propagate along the neighbor filament
channel. We believe that the activation and partial eruption of
the filaments bring the field lines of flux ropes containing them
closer to the null point and trigger the magnetic reconnection
between them and the fan-spine magnetic configuration. As a
result, the hot plasma moves along the outer spine line toward the
remote point. Utilizing the present observations, for the first
time we have discussed how two different-chirality filament channels 
can interact and show interrelation.
\end{abstract}




\section{Introduction}
\label{sec1}

Solar filaments/prominences are characterized as cool and dense
structures that lie above the solar surface in the hot corona
\citep{Labrosse10,Mackay10}. Filaments exist in the magnetic dips
within magnetic configurations known as filament channels or flux
ropes \citep{Aulanier02,Liu12}. Filaments and/or associated
channels sometimes interact and show interesting dynamics in the
chromosphere and low corona
\citep{Uralov02,Schmieder04,Su07,Bone09,Kumar10,Chandra11,Li12,Liu10,Filippov11,Jiang13,Jiang14,Joshi14b}.
During interaction under some specific conditions, these magnetic
structures can  reconnect and change their foot point connectivity.
They can also merge to form one common filament. The first type is
known as 'slingshot' reconnection and the second type is 'merging'
(see paper by \cite{Linton01,Linton05}). Besides, large-scale
flux-rope interaction/merging in the outer corona has also been
observed in the form of CME-CME interaction and their merging
\citep{Gopalswmay01,Joshi13b}.

Some observational studies suggest the observational evidence of slingshot
magnetic reconnection between filaments
\citep{Kumar10,Chandra11,Filippov11,Jiang13}. \cite{Kumar10} and
\cite{Chandra11} first reported the interaction, reconnection, and
footpoint connectivity change between two nearby filaments using
H$\alpha$ observations on 2003 November 20. Later on,
\cite{Filippov11} reported a few observational cases showing pairs
of large filaments joining and exchanging their halves. More
recently, \cite{Jiang13} reported another observational evidence
of partial slingshot reconnection during interaction of two
filaments on 2011 December 3.

There are relatively few numerical simulations, which have been
performed for the slingshot reconnection between flux ropes
\citep{Linton01,Linton05,Torok11a}. \cite{Linton01} and
\cite{Linton05} presented different types of flux rope interaction
using three dimensional magnetohydrodynamic (MHD) simulations for
convection zone conditions. Later on, \cite{Torok11a} simulated
the 2003 November 20 filament interaction using a three
dimensional zero $\beta$ MHD model for coronal conditions and
interpreted it in terms of 'slingshot' reconnection between two
magnetic flux ropes. 

Merging of two filament channels after dynamic interactions have
also been observed by \cite{Schmieder04,Bone09,Jiang14,Joshi14b}.
\cite{Schmieder04} found  evidence of merging of two segments with
dextral chiralities to form a long dextral filament. Recently,
\cite{Jiang14} reported the interaction and merging of two
sinistral filaments on 2001 December 6 and found that they form a
new long magnetic channel. More recently, \cite{Joshi14b} reported
an interesting dynamic event of merging of two filament channels
and formation of a long compound flux rope on 2014 January 1. On
the basis of numerical simulations, some authors discussed various
conditions necessary for interaction/merging of filaments.
\cite{DeVore05} and \cite{Aulanier06} modeled filaments as differentially
sheared arcades and found that two filaments, occupying a single
polarity inversion line (PIL) in a bipolar large-scale magnetic
configuration, easily merged if their chiralities were identical
and axial magnetic fields were aligned. This is in accordance with
empirical rules for filament interaction found by
\cite{Martin94} and \cite{Schmieder04}. However, in a quadrupolar
configuration the situation is more complex and ambiguous
\citep{DeVore05,Linton06,Romano11}. 

\cite{Linton01} and \cite{Linton05} analyzed numerically the reconnection of two twisted flux tubes contacting at different angles. The result of interaction depends on the twist handedness of the tubes and the angle between their axial magnetic fields. A pair of oppositely twisted flux tubes shows a bounce interaction, if their axial magnetic fields are parallel, and slingshot reconnection, if their axial magnetic fields are anti-parallel or perpendicular. \cite{Linton01} and \cite{Linton05} considered isolated flux tubes without a surrounding magnetic field. But flux ropes containing filaments are not isolated flux tubes. They are imbedded into coronal magnetic fields created mostly by photospheric sources and follow basically photospheric PILs. For example, observations of filaments crossing each other at different heights are very rare. Oppositely twisted flux ropes with antiparallel axial magnetic fields need the presence of an additional PIL between them to be in equilibrium in the coronal magnetic field. Therefore, the slingshot interaction of filaments of different chiralities may be different.

In this paper, we present observations 
of interaction of two adjacent filament channels of different chirality 
associated with two adjacent PILs within a fan-spine configuration. 
This kind of dynamic interaction has not been discussed in detail before. 
We discuss the filament-channel interaction dynamics, 
the probable magnetic reconnection at a null point above
them, different helical motions, and the apparent exchange of
heated plasma between different filament channels. The structure
of the paper is as follows: Section~\ref{sec2} deals with the
description of the observational data set used in the paper.
Morphology and magnetic structure of the filaments are discussed
in Section~\ref{sec3}. Different events of interactions and
associated plasma dynamics are described in Section~\ref{sec4}. In
Section~\ref{sec5}, we present an interpretation of the observed
phenomenon in the light of filament flux-rope models and
discussion. Main results and conclusions are listed in
Section~\ref{sec6}.


\section{Observations}
\label{sec2}

We used Big Bear Solar Observatory (BBSO) and {\it National Solar Observatory} (NSO)/ 
{\it Global Oscillation Network Group} (GONG) H$\alpha$
observations for the present study. The BBSO high resolution 
images are collected from archive \url{http://www.bbso.njit.edu/} 
are used to investigate the chiralities of the filaments. 
The GONG data are available in the
data archive at \url{http://halpha.nso.edu/archive.html} with
full-disk images in 6563 \AA~line. The images have spatial
resolution of 1$"$ and a cadence of around 1 min \citep{Har11}.
The GONG H$\alpha$ observations are used to get the information about
the filament activation and partial eruption dynamics. We also
used data of the {\it Atmospheric Imaging Assembly} (AIA;
\cite{Lem12}) instrument on board the {\it Solar Dynamics
Observatory} (SDO). It observes the full disk of the Sun in
ultra-violet (UV) and extreme ultra-violet (EUV) wavelengths with
a minimum cadence of 12 s and a pixel size of 0.6$"$. We used AIA
images in 304 and 171, 193, 131 and 94 \AA\ wavelength channels. The line-of-sight
(LOS) photospheric magnetic field data are obtained by the {\it
Helioseismic and Magnetic Imager} (HMI; \cite{Schou12}), with a
spatial resolution of 1$"$ and a minimum cadence of 45 s. It is
also an instrument on board  {\it SDO}.

\section{Morphological and Magnetic Structure of the Filaments}
\label{sec3}

\begin{figure}
\vspace*{-3cm}
\centerline{
    \hspace*{0.0\textwidth}
    \includegraphics[width=2\textwidth,clip=]{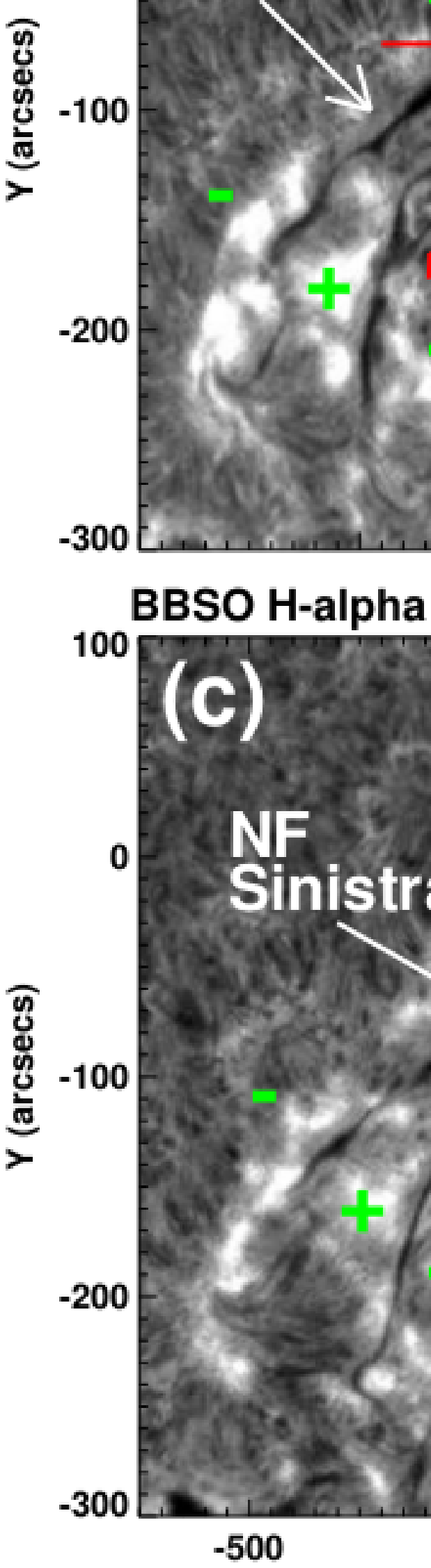}
    }
\vspace*{-2.2cm} 
\caption{BBSO H$\alpha$ images at $\sim$18:46 on 
2014 April 15 (a) and $\sim$17:01 UT on 2014 April 16 (c), showing the 
presence of northern (NF) and southern (SF) filaments. (b) 
\textit{SDO}/AIA 304 \AA\ images at $\sim$18:46 UT on 2014 April 15. Northern filament channel (NFC) and southern filament channel (SFC) are shown in panel (b). 
The \textit{SDO}/HMI line-of-sight magnetogram (d) with overplotted
filament spines shown by red (NF) and orange (SF) colors. 
The filament spines are tracked from the H$\alpha$ image shown in panel (a).}
\label{fig1}
\end{figure}

Figures~\ref{fig1}(a) and \ref{fig1}(c) represent the BBSO H$\alpha$
images at $\sim$18:46 UT on 2014 April 15 and $\sim$17:01 UT on 
2014 April 16, respectively. Figure~\ref{fig1}(b) shows the 
\textit{SDO}/AIA 304 \AA\ images at $\sim$18:46 UT
on 2014 April 15. These images show the filaments and filament
channels about 3-4 days before the first interaction that starts on
2014 April 18. Two dark filaments, named as the northern filament
(NF) and southern filament (SF), are clearly seen  in the
H$\alpha$ image (Figures~\ref{fig1}(a) and (c)). The extended filament
channels of both filaments can be seen in the \textit{SDO}/AIA 304
\AA\ EUV image (Figure~\ref{fig1}(b)). Figure~\ref{fig1}(d) shows
the \textit{SDO}/HMI LOS magnetogram at 18:46:19 UT on 2014 April
15. To find out the filament positions and approximate endpoint
locations, we tracked the filament spines from the H$\alpha$ image
(see Figure~\ref{fig1}(a)) and overplotted them in the LOS
magnetogram (Figure~\ref{fig1}(d)). NF/SF axis are shown by the
red/orange colors, respectively in Figure~\ref{fig1}(d). After comparing
H$\alpha$, EUV 304 \AA\ images, and the LOS magnetogram we
determine that the eastern/western ends of both filaments are
anchored in negative/positive polarity, respectively. Both
filaments are stretched approximately from the south-east to the
north-west along two different PILs.

The handedness or chirality of filaments can be determined using
high-resolution H$\alpha$ images and the position of filament ends
relative to the photospheric LOS magnetic fields.
Figure~\ref{fig1}(d) hints that the eastern ends of both filaments
are anchored in negative polarities, while the western ends are
rooted in positive polarities. In this case, the axial
magnetic field in both filaments is directed from west to east.
Accordingly, NF is sinistral, while SF is dextral because they are
separated by positive polarity.

The H$\alpha$ images in Figures~\ref{fig1}(a) and~\ref{fig1}(c) also show 
that fine threads within NF and SF bodies deviated counterclockwise and 
clockwise from their axes, respectively. The filament barbs are 
left-bearing/ right-bearing for the NF/SF, which corresponds to 
sinistral chirality of NF and dextral chirality of SF 
\citep{Martin94,Martin98a}. Some of the visible barbs are marked 
with the red arrows in Figures~\ref{fig1}(a) and~\ref{fig1}(c).

\begin{figure}
\vspace*{-3cm}
\centerline{
    \hspace*{0.0\textwidth}
    \includegraphics[width=1.8\textwidth,clip=]{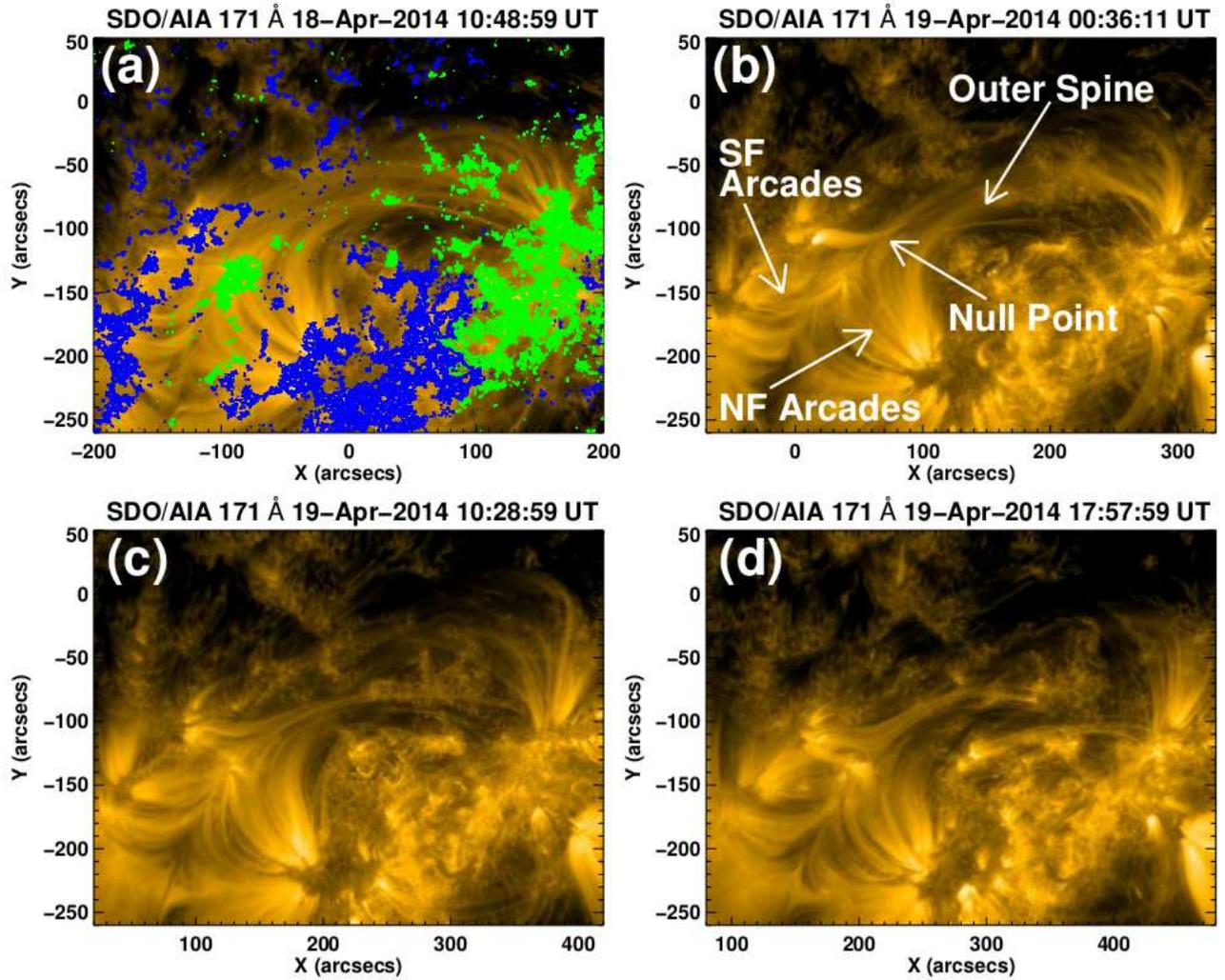}
    }
\vspace*{-2.4cm} 
\caption{\textit{SDO}/AIA 171 \AA\ images on 2014
April 18--19 showing the fan-spine type configuration over two
filament channels. The southern filament (SF) arcades, northern filament (NF) arcades,
the approximate location of null point and the outer spine are marked
in panel (b).}
\label{fig2}
\end{figure}

\begin{figure}
\vspace*{-3cm}
\centerline{
    \hspace*{0.0\textwidth}
    \includegraphics[width=1.6\textwidth,clip=]{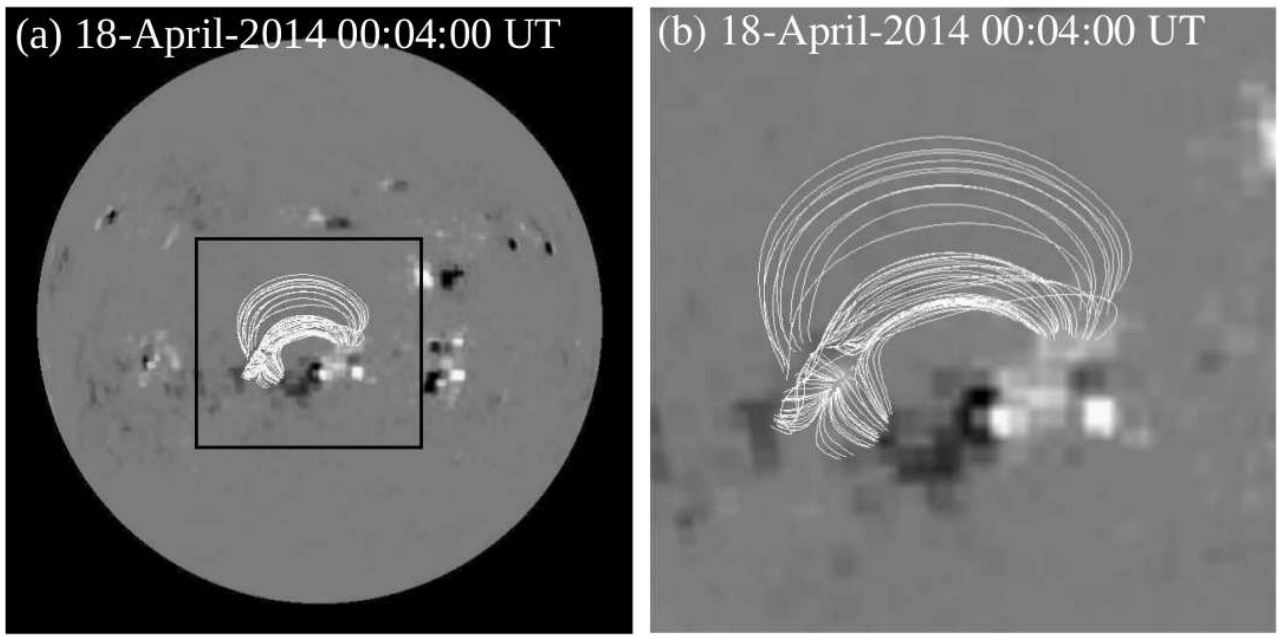}
    }
\vspace*{-3.8cm} 
\caption{(a) PFSS magnetic field extrapolation
showing a full disk view of the coronal magnetic field structure
over the filaments. (b) The zoomed view corresponding to the black
box shown in the panel (a). } 
\label{fig3}
\end{figure}

The magnetic configuration surrounding the filaments can be
deduced from the analysis of \textit{SDO}/AIA 171 \AA\ images
(Figure~\ref{fig2}) and the potential-field source-surface (PFSS)
magnetic-field extrapolation \citep{Schrijver03}
(Figure~\ref{fig3}). Figures~\ref{fig2}(a)--(d) represent
\textit{SDO}/AIA 171 \AA\ images at 10:48:59 UT on 2014 April 18
and at 00:36:11 UT, 10:28:59 UT, and 17:57:59 UT on  2014 April
19, respectively. To compare the coronal loop structure with a LOS
magnetogram, we overplot positive (green) and negative (blue) LOS
magnetic field contours on the AIA 171 \AA\ image
(Figure~\ref{fig2}(a)). We clearly see arcades, connecting the
central positive polarity with negative polarities on both sides,
above the filaments. We also see long loops that connect the
negative polarities to the remote region of positive polarity,
near to the western endpoint of SF. All these images show the
fan-spine configuration over the filaments. A null point is
expected to be above the central positive polarity between the
filaments. The outer spine field line emanates from the null point
but is directed not radially into the outer corona, as it is
usually assumed in fan-spine configurations, but deviates to the
west and touches the photosphere  near the western endpoint of SF
within an area of positive polarity. This configuration is
confirmed by the PFSS magnetic field extrapolation
(Figure~\ref{fig3}). To perform the PFSS extrapolation we used the
PFSS software package available in IDL SolarSoftWare.
Figure~\ref{fig3}(a) shows the full disk magnetogram  with
extrapolated magnetic field lines, while the zoomed region
corresponding to the black box is shown in Figure~\ref{fig3}(b).
The calculated field lines match quit well to the structure of
coronal loops in \textit{SDO}/AIA 171 \AA\ images.

\begin{figure}
\vspace*{-4.5cm}
\centerline{
    \hspace*{0.0\textwidth}
    \includegraphics[width=2.2\textwidth,clip=]{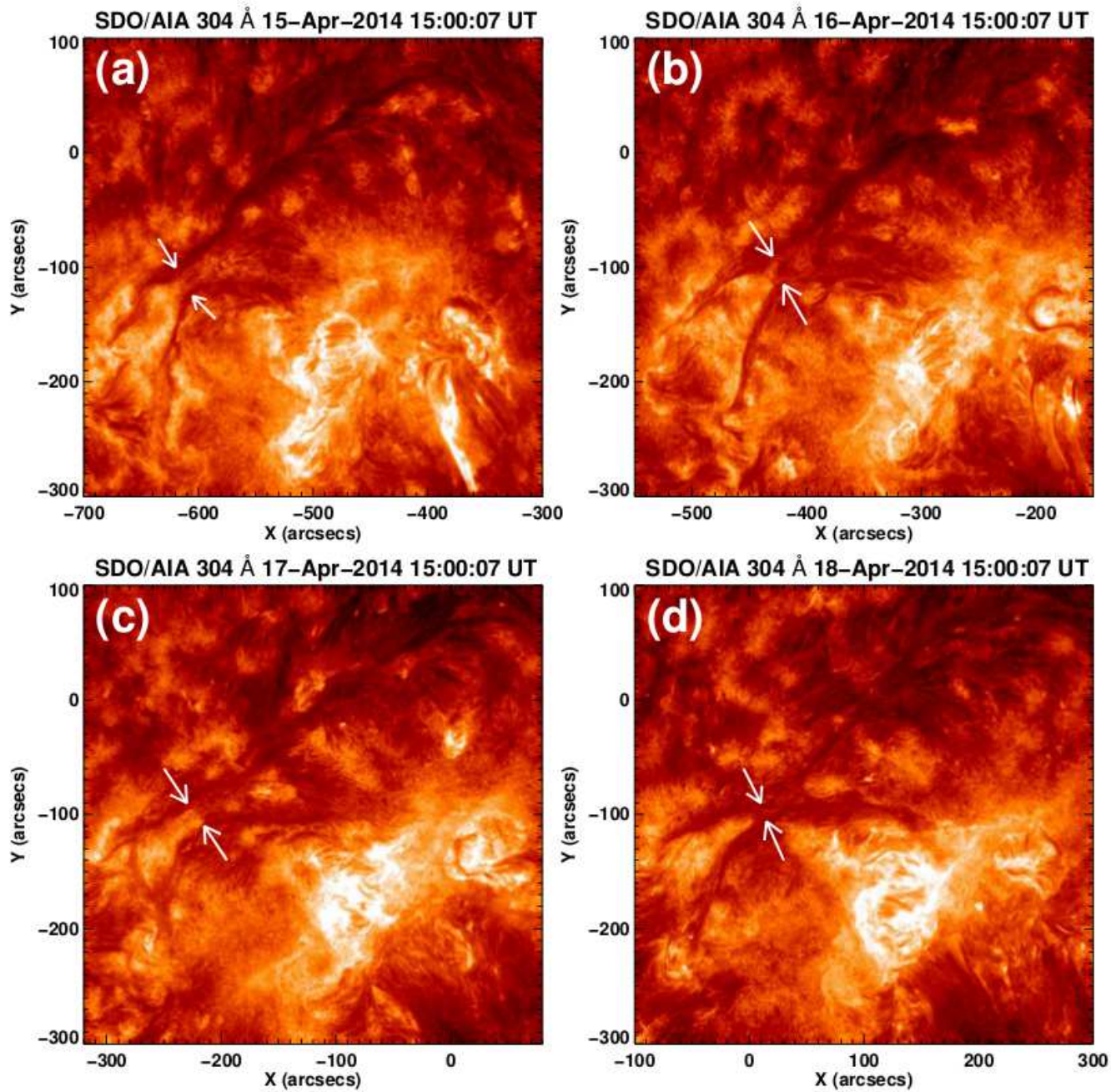}
    }
\vspace*{-2.5cm} 
\caption{Evolution of filament channels during
2014 April 15-18  in \textit{SDO}/AIA 304 \AA\ images. Arrows show
the closing in of filament channels.} 
\label{fig4}
\end{figure}

The filament channels were approaching each other from April 15 to
April 18 as seen in \textit{SDO}/AIA 304 \AA\ images in
Figure~\ref{fig4}.  The filament channels manifest themselves as
long dark structures.  They are quite separated on April 15
(Figure~\ref{fig4}(a)), come slowly closer to each other during
April 16-17 (Figure~\ref{fig4}(b)-(c)), and become very close on
April 18 (Figure~\ref{fig4}(d)). The closing of the filament
channels is marked by the white arrows in all the panels in
Figure~\ref{fig4}.
\section{Dynamic Interactions of the Filament Channels}
\label{sec4}

We observed three events of filament channel interactions during 
2014 April 18--20. In this section we describe the detail observation 
of these interaction dynamics in multiwavelength channels.

\subsection{First Event of Interaction and Associated Dynamics}
\label{sec4.1} 

\begin{figure}
\vspace*{-6cm}
\centerline{
    \hspace*{0.0\textwidth}
    \includegraphics[width=3\textwidth,clip=]{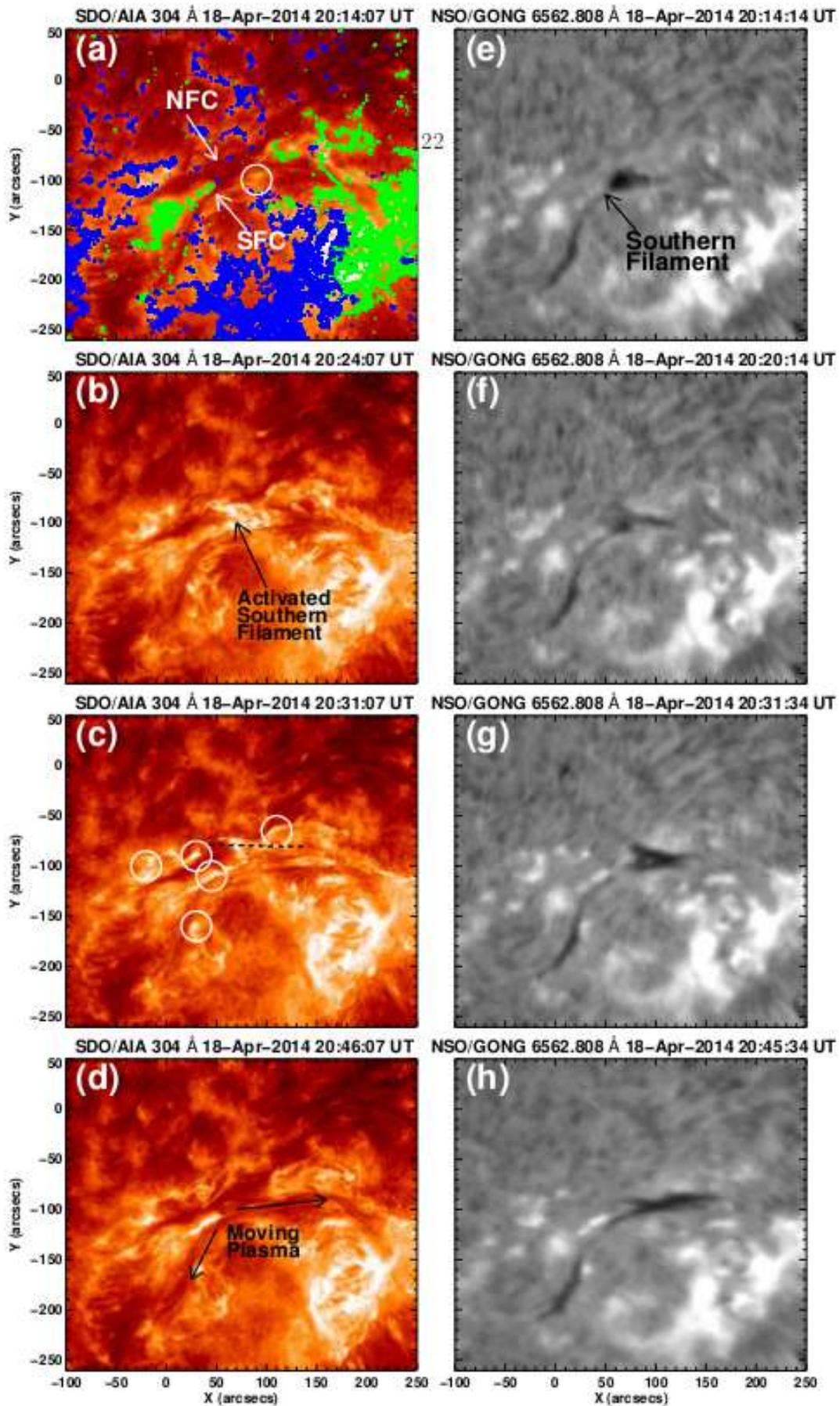}
    }
\vspace*{-3.8cm} 
\caption{\textit{SDO}/AIA 304 \AA\  (left column)
and the NSO/GONG H$\alpha$ (right column) images showing the first
event of interaction/reconnection. Northern filament channel (NFC) and southern filament channel (SFC) are shown in panel (a).} 
\label{fig5}
\end{figure}

The sequence of the images showing the first event
of interaction are represented in Figure~\ref{fig5}. The left
panel shows the \textit{SDO}/AIA 304 \AA\ images ((a)--(d)), while
the right panel show the NSO/GONG H$\alpha$ images ((e)--(h)). In
Figure~\ref{fig5}(a), the \textit{SDO}/AIA 304 \AA\ image is
overplotted with \textit{SDO}/HMI LOS magnetogram contours.
Green/blue contours show the positive/negative polarity regions,
respectively. In Figure~\ref{fig5}(a), we  see two nearby filament
channels. However, at the same time in the H$\alpha$ image only SF
is visible (Figure~\ref{fig5}(e)). The first event of interaction
starts around 20:14 UT on 2014 April 18 with the activation of 
the middle part of SF. The initial activation area is shown by the
small white circle in Figure~\ref{fig5}(a). After the activation,
the middle part of the filament partially erupts towards the
north. The activated filament is seen in Figure~\ref{fig5}(b).
Along with the partial eruption we also see the counterclockwise
rotation of filament threads around the long filament axis, if we
observe it from the east end (see the AIA 304 \AA\ animation
associated with Figure~\ref{fig5}). This counterclockwise rotation
of the threads can be the manifestation of the redistribution of the 
twist along the flux rope due to its expansion and swelling during 
the activation \citep{Parker74}.
Negative helicity of a flux rope corresponds to the dextral chirality of a
filament in flux-rope models.

The partial failed eruption of SF seems to trigger reconnection at
the magnetic null that lie above the filament channels. It is
exhibited by EUV brightenings at several places on either side of
both filament channels simultaneously with the partial eruption
and helical motion. These bright regions are marked by the white
circles in Figure~\ref{fig5}(c). Such brightenings are believed to
appear due to hits of the chromosphere by fast electrons and
heated plasma from the region of reconnection. After the partial
failed eruption the heated as well as cool material of SF moves
along the axis towards the eastern and western ends of the
filament (directions are shown by the arrows in
Figure~\ref{fig5}(d)). The two separate filament channels are
still observed very close to each other at 20:46:07 UT. Apart from
the chromospheric brightenings, no influence of SF activation on
NF was observed.

\begin{figure}
\vspace*{-3cm}
\centerline{
    \hspace*{0.0\textwidth}
    \includegraphics[width=1.8\textwidth,clip=]{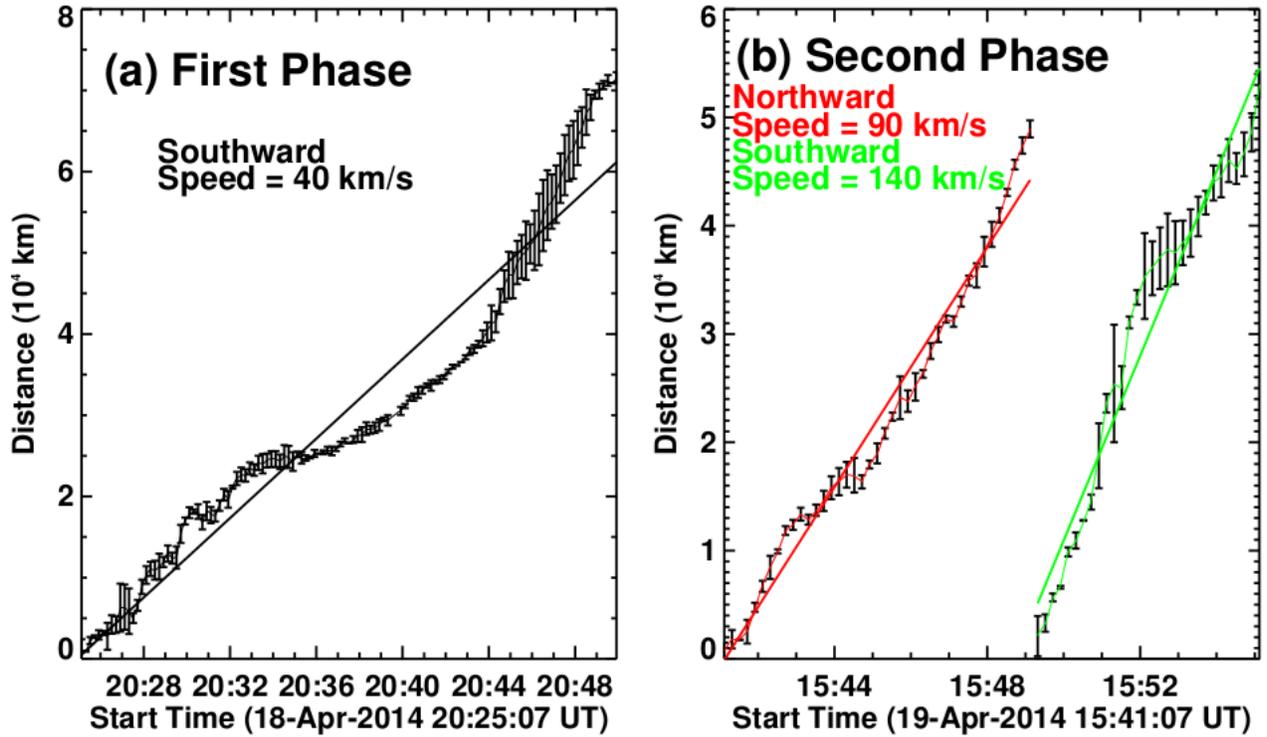}
    }
\vspace*{-4.2cm} 
\caption{(a) Projected distance-time profile of the
plasma flow along the trajectory shown by the dashed black line in
Figure~\ref{fig5}(c) measured using \textit{SDO}/AIA 304 \AA\
images. (b) The same for the trajectories shown by the dashed
white (in red color) and black lines (by green color) in Figure~\ref{fig7}(d). The error bars are
the standard deviations estimated using three repeated
measurements of the same value. The linear fit method has been used to estimate the average speeds.} 
\label{fig6}
\end{figure}

The distance--time plot of the hot plasma movement is presented in
Figure~\ref{fig6}(a). The rough trajectory along which the distance 
measurements was made is shown by dashed black line in
Figure~\ref{fig5}(c). We tracked a bright plasma blob that moves
towards the western end of SF. The east-most point was used as a
reference point for the distance measurements. For more accurate
results, we repeated the measurements three times and the standard
deviations was used as errors. The linear fit to these data points
is used to estimate an average speed. It is evident that the hot
plasma moved with the average speed of $\sim$40 $\rm km~s^{-1}$
between 20:25 and 20:50 UT.
\subsection{Second Event of Interaction and Associated Dynamics}
\label{sec4.2}

\begin{figure}
\vspace*{-6cm}
\centerline{
    \hspace*{0.0\textwidth}
    \includegraphics[width=3\textwidth,clip=]{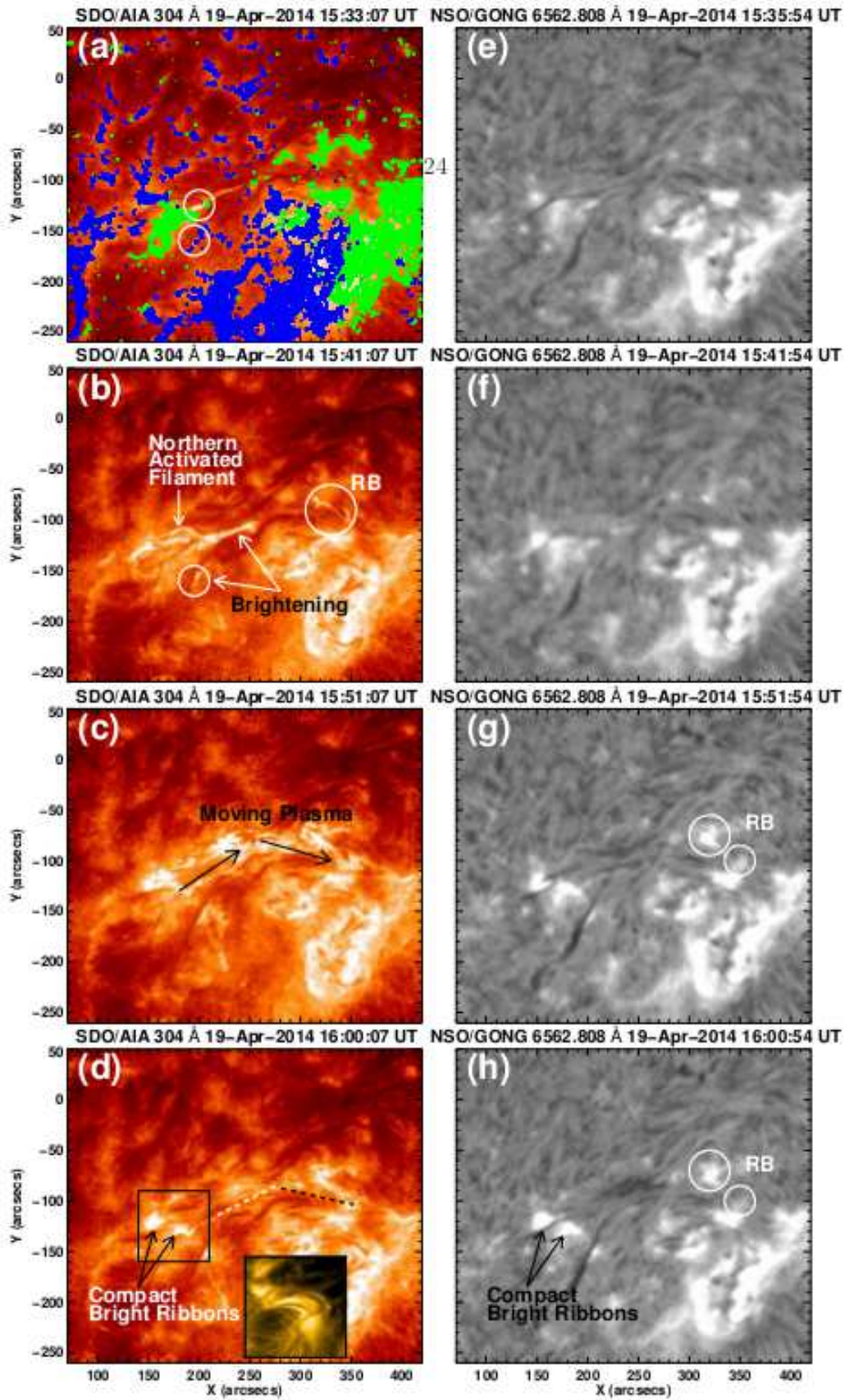}
    }
\vspace*{-4.4cm} 
\caption{\textit{SDO}/AIA 304 \AA\  (left column)
and the NSO/GONG H$\alpha$ (right column) images showing the
second event of interaction/reconnection. The inset image over panel (d) 
is SDO/AIA 171 \AA\ image at $\sim$16:21 UT on 2014 April 19, showing 
the post flare loops joining the two compact bright ribbons. Remote brightening regions are shown by white circles in panels (b), (g) and (h).} 
\label{fig7}
\end{figure}

Figure~\ref{fig7} represents the selected \textit{SDO}/AIA 304
\AA\ and NSO/GONG H$\alpha$ images showing the second event of
interaction and associated dynamics. Second event of interaction
started at $\sim$15:33 UT on 2014 April 19 with a compact
brightening near the place of the closest approaching  of the
filament channels, just between them. Another small brightening
appeared on the southern side of SF. The locations of these
compact brightenings are marked by the white circles in
Figure~\ref{fig7}(a). The activation of NF started at $\sim$15:37
UT near its eastern end (Figure~\ref{fig7}(b)). At the same time,
we also see a remote brightening (RB) on the west. The location of
RB region is shown in Figure~\ref{fig7}(b) with the white circle.
Thereafter, the bright features propagate from the eastern end of
NF to its middle part. The pattern of filament bright and dark
threads looks like the upper part of a right-handed helix, which
is consistent with the sinistral chirality of NF. What is most
surprising, after reaching the place of the closest approaching of
the filament channels the heated plasma propagates not to the
north-west along the axis of the NF channel but to the west and
south-west nearly along the axis of SF. At first glance, one might
fancy that the eastern part of NF and the western part of SF form
a joint magnetic structure allowing plasma to move easily from the
eastern end of NF to the western end of SF. However, it is very
doubtful if they can form such a structure because their chirality
and helicity are opposite. We will discuss the problem in more
detail in Section~\ref{sec5}.

\begin{figure}
\vspace*{-4.5cm}
\centerline{
    \hspace*{0.0\textwidth}
    \includegraphics[width=1.8\textwidth,clip=]{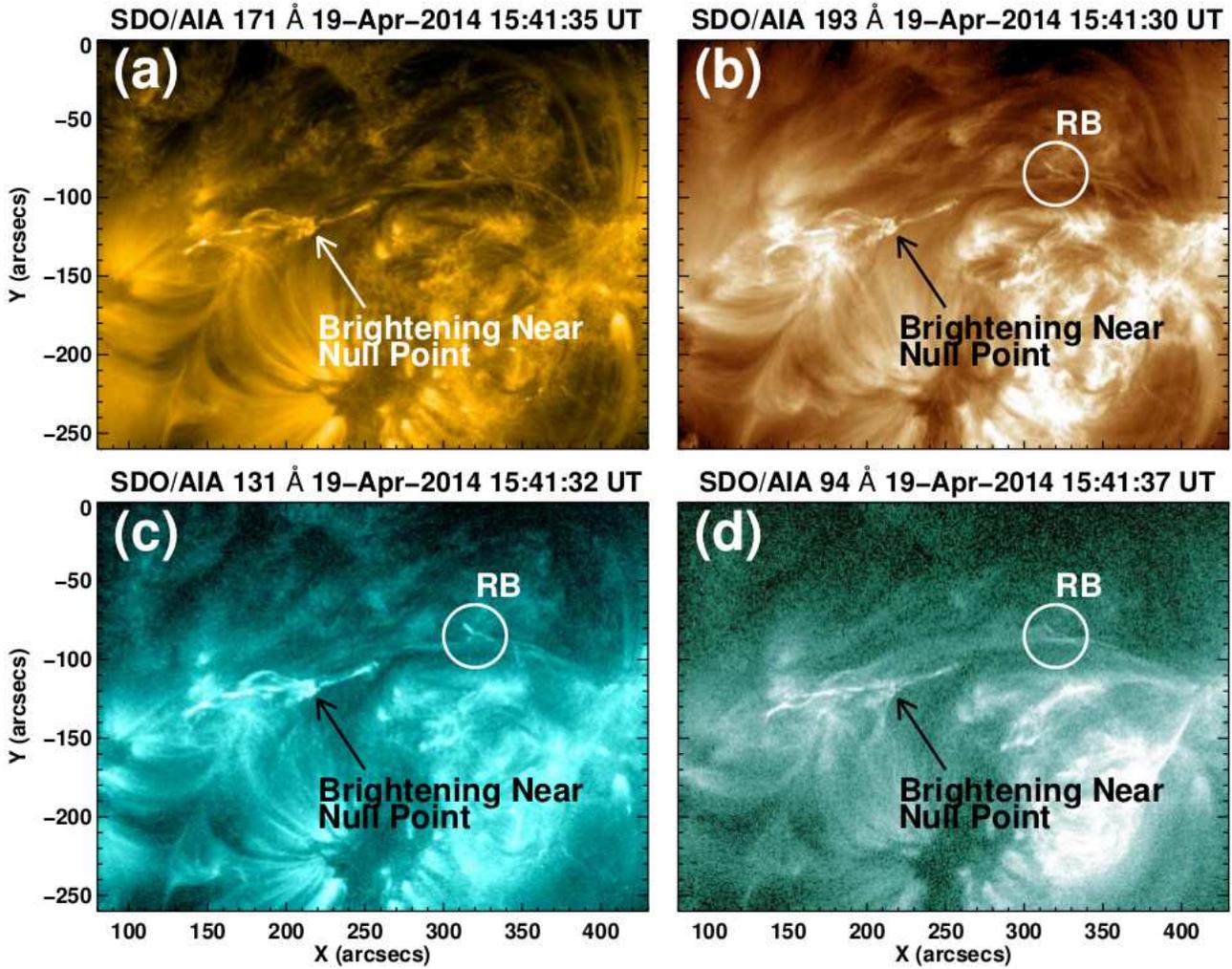}
    }
\vspace*{-2.6cm} 
\caption{\textit{SDO}/AIA 171 (a), 193 (b), 131 (c), 94 (d) \AA\ images at $\sim$15:41 UT 
on 2014 April 19 showing the brightening near the magnetic null point and the 
appearance of remote brightening (RB). The RB area are marked with a white circle 
in panels (b), (c) and (d).} 
\label{fig8}
\end{figure}

Signature of brightening around the magnetic null can also be seen in other AIA 
channels. Figure~\ref{fig8} shows the \textit{SDO}/AIA 171, 193, 131 and 94 
\AA\ image at $\sim$15:41 UT just after the partial filament eruption.
We can see the brightening near the null point, which can be understood
due to the magnetic reconnection there. The remote brightening signature 
is also visible in these channels.

The heated plasma first moves towards northwest direction with a
speed of  $\sim$90 $\rm km~s^{-1}$ and then towards southwest
direction towards the western end of SF with a speed of $\sim$140
$\rm km~s^{-1}$. The kinematics of the plasma flows is shown in
Figure~\ref{fig6}(b). It represents the distance--time profiles of
the plasma flows towards the north-west (red curve) and the
south-west (green curve). The trajectories along which the
northwest directed (white dashed line) and southwest directed
(black dashed line) displacement measurements have been performed
are shown in Figure~\ref{fig7}(d). Two compact bright areas
appeared on both sides of NF near its eastern end closer to the
end of the second event (Figures~\ref{fig7}(d), (g) and (h)). They
can be considered as the compact ribbons formed by the partial
eruption of the NF and the associated flare. The overplotted \textit{SDO}/AIA 171 \AA\ image 
at around $\sim$16:21 UT shows the loop like structures joining 
the two bright ribbons and can be considered as the post flare loops. 

\subsection{Third Event of Interaction and Associated Dynamics}
\label{sec4.3}

\begin{figure}
\vspace*{-6cm}
\centerline{
    \hspace*{0.0\textwidth}
    \includegraphics[width=3\textwidth,clip=]{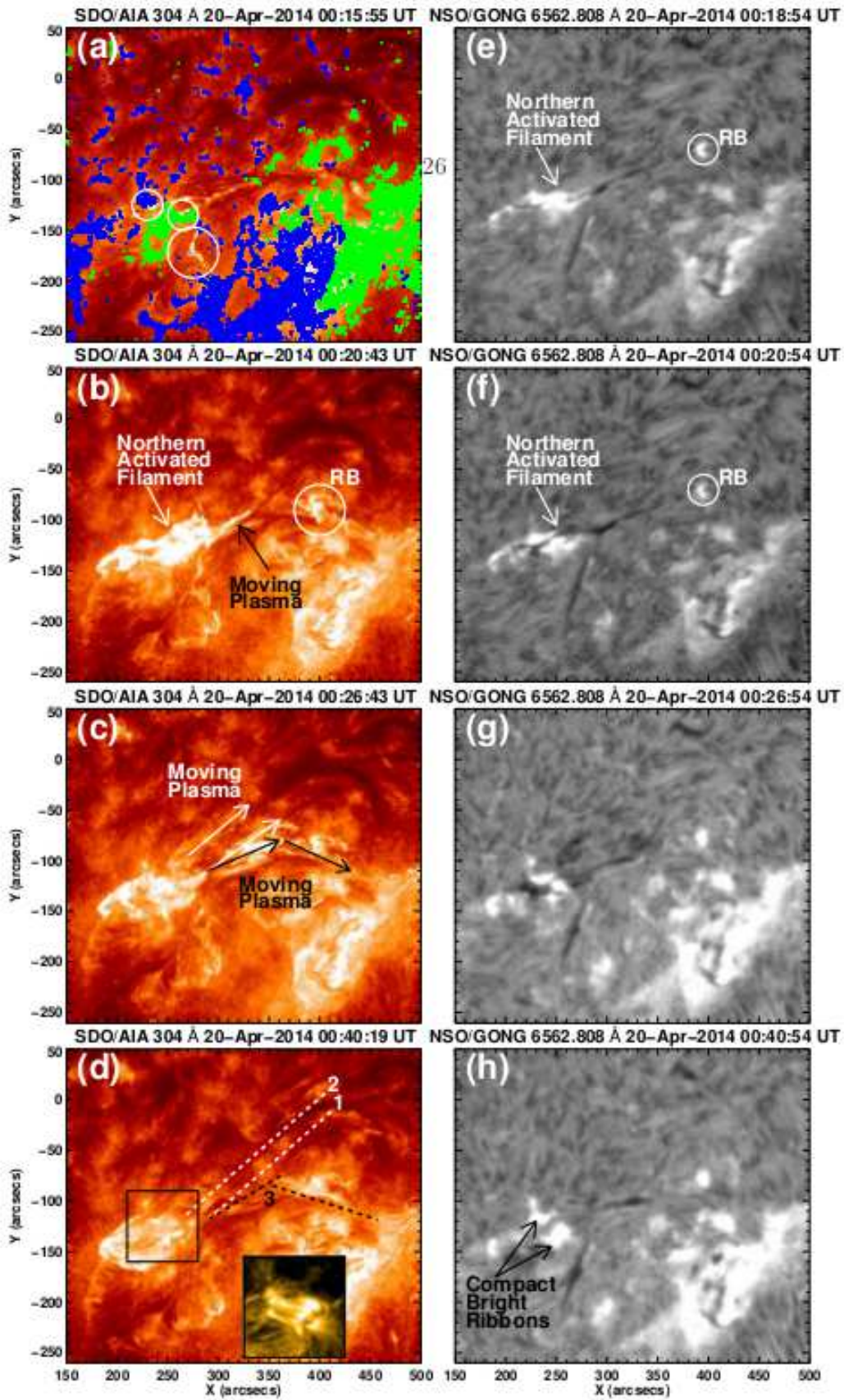}
    }
\vspace*{-4.3cm} 
\caption{\textit{SDO}/AIA 304 \AA\  (left column)
and the NSO/GONG H$\alpha$ (right column) images showing the third
event of interaction/reconnection. The inset image over panel (d) 
is SDO/AIA 171 \AA\ image at $\sim$00:40 UT on 2014 April 20, showing 
the post flare loops joining the two compact bright ribbons (panels g and h). Remote brightening regions are shown by white circles in panels (b), (e) and (f).} 
\label{fig9}
\end{figure}

\begin{figure}
\vspace*{-4.5cm}
\centerline{
    \hspace*{0.0\textwidth}
    \includegraphics[width=2\textwidth,clip=]{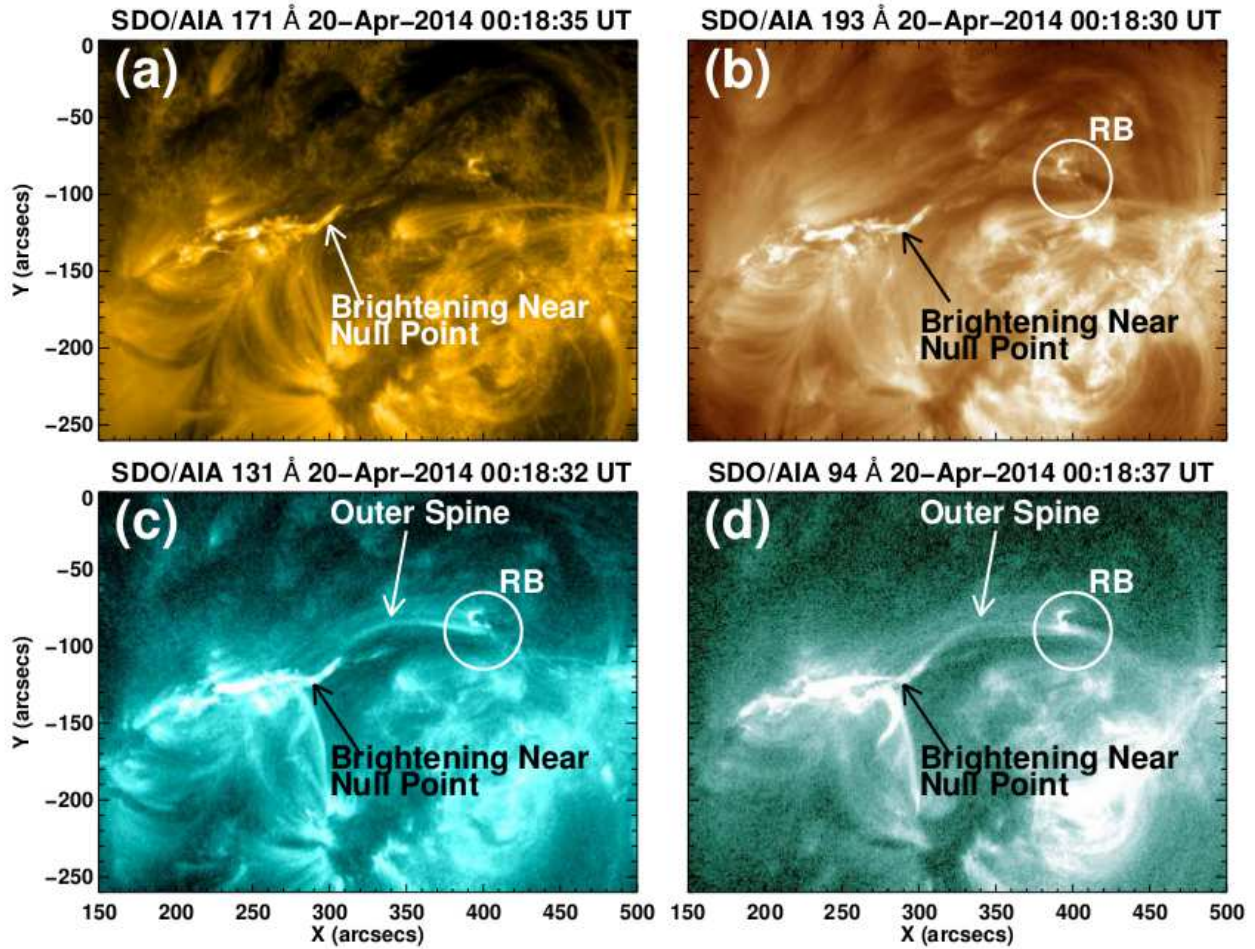}
    }
\vspace*{-3.6cm} 
\caption{\textit{SDO}/AIA 171 (a), 193 (b), 131 (c), 94 (d) \AA\ images at $\sim$00:18 UT 
on 2014 April 20 showing the brightening near the magnetic null point 
and the appearance of remote brightening (RB) and outer spine. The RB area are 
shown by white circle in panels (b)--(d), while the outer
spine are represented in panels (c) and (d).} 
\label{fig10}
\end{figure}

The third event in many features is similar to the second event of
interaction. Figure~\ref{fig9} show the interaction dynamics in
\textit{SDO}/AIA 304 \AA\ and NSO/GONO H$\alpha$ observations.
It starts at $\sim$00:14 UT on 2014 April 20 with
brightenings near the place of the closest approaching  of the
filament channels and the southern side of SF. Immediately after
that at $\sim$00:15 UT we observe an activation and failed
eruption of the eastern part of NF with formation of two bright
ribbons on both sides of it. These ribbons can be formed as a result of reconnection between the legs of surrounding arcades during the partial eruption of the NF inside the northern arcades of the fan-spine structure. Soon after the activation, a remote
brightening appears near the western end of SF at nearly the same
place as in the second event (Figure~\ref{fig9}(b), (e) and (f)).
We also observed the brightening at the magnetic null point just 
after the partial eruption of NF in All the EUV channels 
(Figures~\ref{fig9}(b) and~\ref{fig10}). Figure~\ref{fig10} represents 
\textit{SDO}/AIA 171, 193, 131 and 94 \AA\ images at $\sim$00:18 UT 
on 2014 April 20 also showing the brightening at the null. This brightening 
is due to some magnetic reconnection at the null point. The outer 
spine lines and the remote brightening can be seen in the hotter 
AIA channels (Figures~\ref{fig10}(c) and (d)).

Heated plasma of the eastern part of NF forms a wide bright
helical structure with intensive internal motions. Some part of
the hot plasma moves from the middle of NF to the north-west along
its axis to the western end, while a fraction of  bright material
moves to the western end of SF along the curved path nearly the
same as in the second event. Different directions of plasma
flows are shown by the arrows in Figure~\ref{fig9}(c).

Several long threads as a whole shift from the northern side of
the NF channel to the southern side. This movement corresponds to
clockwise rotation of a right-hand helix around its axis, as seen
from the east, if the threads belong to its upper part and reveals
untwisting of the helix. The eastern part of the helix looks more
twisted, with threads more transversal to the axis. At the ending
phase of the event, there are many blobs moving along the threads
to the eastern end of NF. Their rotation (counterclockwise) is
opposite to the rotation of the whole threads in the middle part
of the helix because they presumably move along the upper part of
the right-hand helix to its eastern end.

\begin{figure}
\vspace*{-3cm}
\centerline{
    \hspace*{0.0\textwidth}
    \includegraphics[width=2.0\textwidth,clip=]{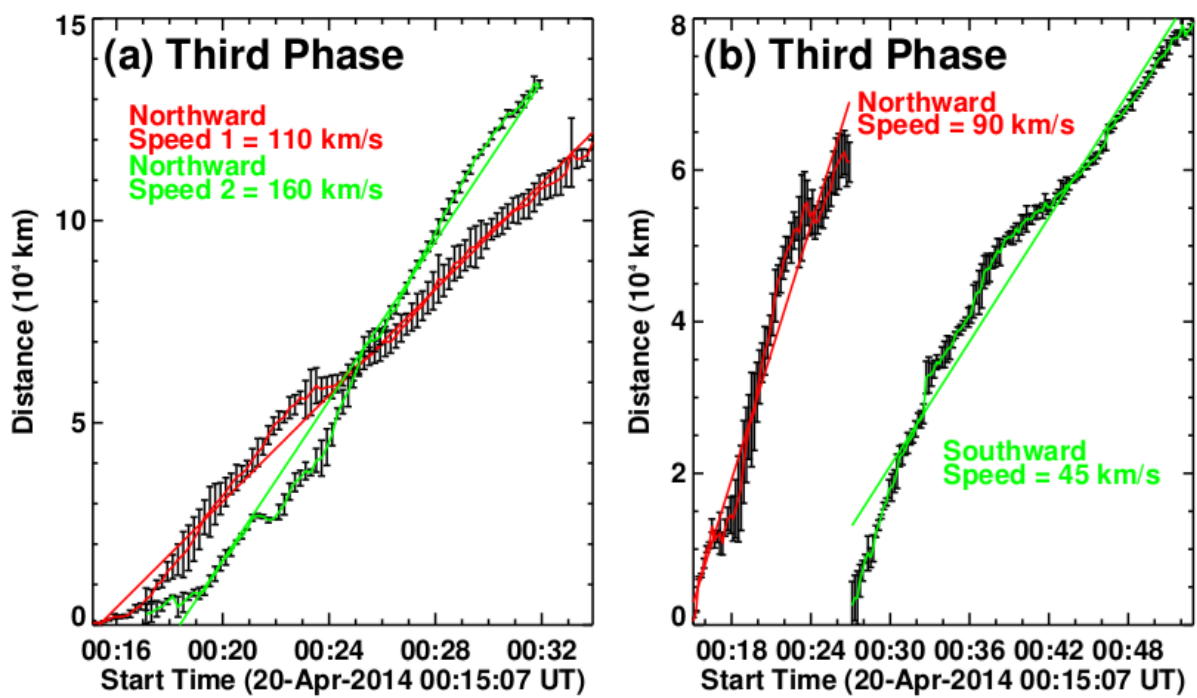}
    }
\vspace*{-6cm} 
\caption{(a) Projected distance--time profiles of the 
plasma flows along the trajectories 1 (red) and 2 (green). (b) The same for the northward (red) and southward (green) plasma flows along the trajectorie 3. These measurements have been made using \textit{SDO}/AIA 304 \AA\ images. The rough trajectories 1, 2 (by dotted white lines), and 3 (by dotted black lines) are shown in Figure~\ref{fig9}(d). The error bars are the standard deviations estimated using three repeated measurements of the same region. The linear fit method has been used to estimate the average speeds.}
\label{fig11}
\end{figure}

The distance--time profiles of these plasma motions are
represented in Figure~\ref{fig11}. Figure~\ref{fig11}(a) shows the
results for plasma moving to the north-west along the NF channel.
We measured the profiles along two different trajectories shown by
white dashed lines in Figure~\ref{fig9}(d). Heated plasma moves
with average speeds of $\sim$110 and $\sim$160 $\rm km~s^{-1}$
along trajectories 1 and 2, respectively. Figure~\ref{fig11}(b)
represents the profiles of plasma motion along the SF channel
towards its western end. Since plasma moves along the curved path,
we measured two profiles along straight lines, one for the
northwestward motion and another for the southwestward motion. The
speeds are $\sim$90 and $\sim$45 $\rm km~s^{-1}$, respectively.

\section{Interpretation and Discussion}
\label{sec5}

Two filaments gradually approach each other in their middle parts
during their passage through the solar disk on 2014 April. We
specify chiralities of the filament to be opposite. The NF fine
structure definitely reveals the sinistral chirality, which is in
accordance with the general hemispheric rule for the southern
hemisphere. The chirality of SF is evident to be dextral for 
many reasons despite the violation of the
hemispheric rule (see Section~\ref{sec3} and Figure~\ref{fig1} 
for more details). During several episodes from April 18 to April
20, the filaments show an activation and formation of a temporal
structure that joins them into a united system. It looks puzzling
because usually filaments with parallel axial magnetic fields and opposite chirality do not merge or
reconnect with the formation of new stable or two different filaments 
from their halves.

In our interpretation of these observations we follow the
flux-rope model \citep[e.g.,][]{Canou10,Guo10,Joshi14,Filippov15} of filaments considering filament plasma
accumulated in lower parts of helical flux tubes. Dextral
filaments are contained within left-handed helices, while
sinistral filaments fill right-handed ones. In our case, the axial
magnetic fields of the two flux ropes are parallel but the
azimuthal fields have different sense of rotation. Therefore, when
these two flux ropes come close together side-by-side, both their
axial and azimuthal field components have the same directions and
cannot reconnect. 

Two flux ropes of similar helicity either show 
merging or slingshot reconnection during their interaction
\citep{Linton01,Torok11a}. However, two flux ropes of different
helicity with parallel axial magnetic fields can only undergo a bounce interaction \citep{Linton01}
when they are brought together. They repulse from each other and
cannot reconnect and form the joint structure. In our case, it is
clear that SF/NF have dextral/sinistral chiralities, respectively with parallel axial magnetic fields. Therefore,
the associated flux ropes should have different signs of twist,
which is the condition for the bounce interaction.

We believe that although the events look like interaction of two
filament channels, the most important interaction occurs between a
flux rope and the surrounding coronal magnetic field of special
structure. We clearly observe a fan-spine configuration of coronal
loops over the filament channels, with a presumed null point above
them (see Figure~\ref{fig2}). The PFSS magnetic field extrapolation 
confirms the existence of the fan-spine magnetic configuration (see Figure~\ref{fig3}).

\begin{figure}
\vspace*{-6cm}
\centerline{
    \hspace*{0.0\textwidth}
    \includegraphics[width=2.8\textwidth,clip=]{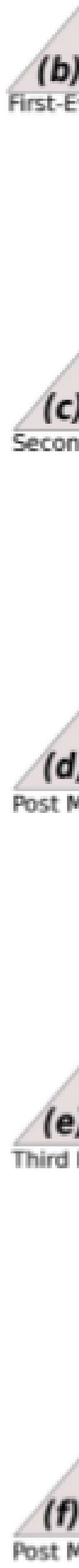}
    }
\vspace*{-2.9cm} 
\caption{Schematic representation of the change in magnetic configuration and the 
reconnections senario in the {3D} disk view (left column) and projected view of 3D field lines in a 2D plane (right column). 
The northern (NF) and southern (SF) filaments are shown by the red and orange colors, respectively. 
Reconnection regions are marked by the pink star, while the reconnected field 
lines are shown by the dotted black lines. The blue overlying arcades over NF in panels (c), (d), (e) and (f) are not in the plane of reconnection.}
\label{fig12}
\end{figure}

The initial magnetic field-line distribution and subsequent plasma
dynamics are shown in the schematic representation in
Figure~\ref{fig12}. 
The coronal structure is similar to a
"pseudostreamer" \citep{Wang07,Rachmeler14} with two flux ropes at
the base. The left column represents the 3D disk view (Figure~\ref{fig12}(a)), while the right column shows the projected view of selected 3D field lines (Figure~\ref{fig12}(g)). 
However, in contrast to the "pseudostreamer" the outer
spine field line emanating from the null point is not directed
radially into the outer corona, but deviates to the west and
touches the photosphere near to the endpoint of SF within an area
of positive polarity (see Figures~\ref{fig2} and~\ref{fig3}). 
In projection on the disk, the outer spine
field line runs nearly parallel to the SF axis, so the plasma motion 
along the spine can easily be mixed up with the motion along the
SF axis. We believe it is most probable in the observations of the
filament interaction in our case.

In the first event the case is looking simple, i.e., reconnection between the inner green and the outer blue line (Figures~\ref{fig12}(b),~\ref{fig12}(h) and~\ref{fig5}). However in the second and third case the senario is little complex. 
There is no anti-parallel field lines belonging to flux ropes that
contain SF and NF. But if each of the flux ropes approaches the
null point, its azimuthal field can reconnect with outer field 
lines of the opposite lobe, i.e., circular 
field lines of the red flux rope can reconnect with the outer 
green line. This case seems to take place in the second (Figures~\ref{fig12}(c),~\ref{fig12}(i), and~\ref{fig7}) and third (Figures~\ref{fig12}(e),~\ref{fig12}(k), and~\ref{fig9}) events of the filament interaction.  The
locations of reconnections are shown by pink stars in
Figure~\ref{fig12}. Due to the projected view of 3D field lines in a 2D plan, the reconnected field lines are appear as a single line in panels (j) and (l) of Figure~\ref{fig12}. But actually it represents the two different sets to field lines as shown in 3D view (Figures~\ref{fig12}(d) and~\ref{fig12}(f)).

After the reconnection some amount of heated plasma confined
previously within the flux rope is able to propagate along the
field lines of the surrounding magnetic configuration. In
particular, it can move along the spine and this motion mimics
the movement along the SF axis. Penetration of the flux-rope
plasma into the outer structure can be illustrated by a simple 2-D
model.

Let us consider the coronal magnetic field with fan-spine
structure as a sum of a vertical homogeneous magnetic field $B_0$
and a vertical 2-D dipole  located at $x = 0$, $z = z_d$ with the
dipole moment $M$. If $y$ is the axis of translational symmetry,
$x$ is the horizontal axis and $z$ is the vertical axis with the
origin at the photospheric level, the external field is described
by $y$-component of vector potential $\bf A$

\begin{equation}
 A_y^e = B_0 x + \frac{M x}{x^2 + (z - z_d)^2} .
\end{equation}

We put into this field a flux rope in the simplest form of a
straight linear current along the $y$-axis. According to the
boundary condition for the coronal current $I$ on the photosphere,
its vector potential can be written as
\citep{VanTend78,Molodenskii87,Filippov01}

\begin{equation}
 A_y^I = \frac{I}{c}[\ln \left((x - x_0)^2 + (z + z_0)^2 \right) -
 \ln \left((x - x_0)^2 + (z - z_0)^2 \right)] ,
\end{equation}
where $x_0$ and $z_0$ are the coordinates of the coronal current.
Neglecting the weight of the flux rope its equilibrium position
$(x_0, z_0)$ is defined by the equations:

\begin{equation}
B_0 - M \frac{ x_0^2 - (z_0 - z_d)^2}{\left(x_0^2 +
 (z_0 - z_d)^2\right)^2} = 0,
\end{equation}

\begin{equation}
 \frac{I}{c z_0} - M \frac{2 x_0 (z_0 - z_d)}{\left(x_0^2 +
 (z_0 - z_d)^2\right)^2} = 0.
\end{equation}

\begin{figure}
\centerline{
    \hspace*{0.0\textwidth}
    \includegraphics[width=1.7\textwidth,clip=]{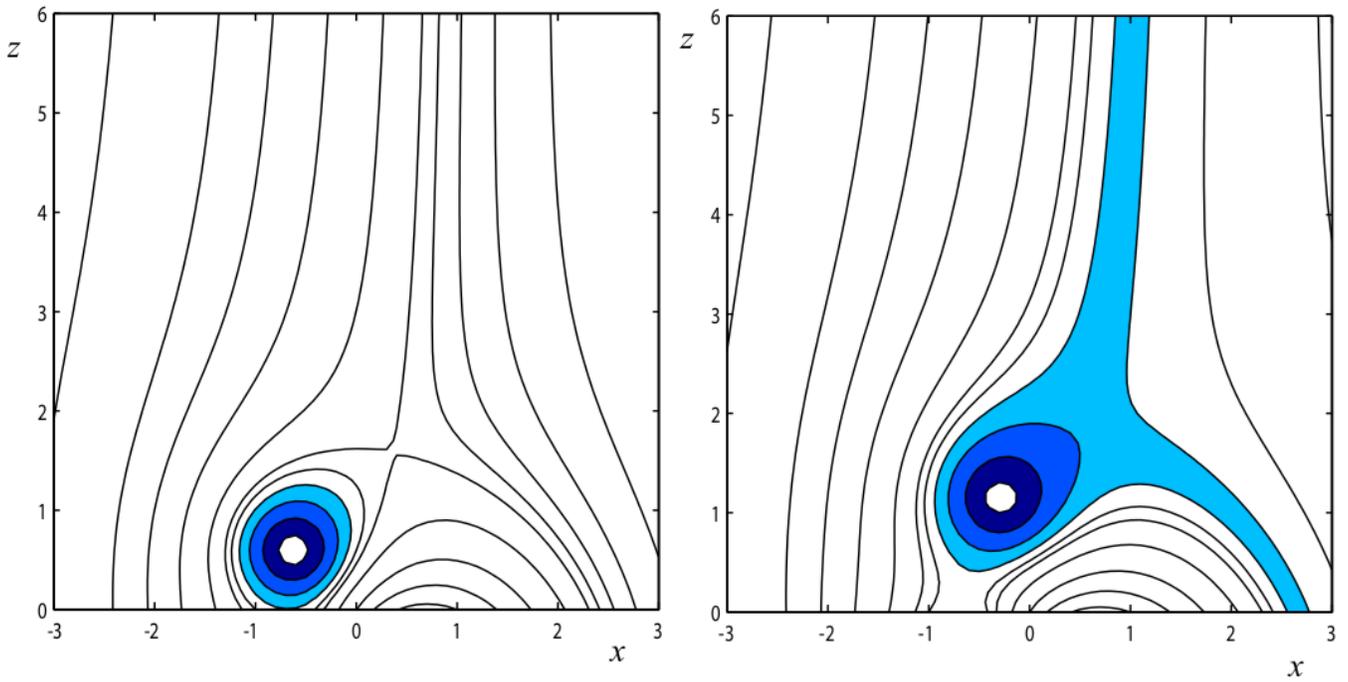}
    }
\vspace*{-4.8cm} 
\caption{2-D model of the flux-rope magnetic field
reconnection at the null point. Black lines represent the magnetic
field lines while the different tints of the  blue color show the
confined plasma.} 
\label{fig13}
\end{figure}

Figure~\ref{fig13}(left) shows field lines described by Equations
(1) and (2) for dimensionless parameters $B_0 = 1, M = -4, z_d =
-1$, and the value of $I/c$ being close to the critical value
of the current $I_c$ over which the stable equilibrium is
impossible \citep{Molodenskii87,Filippov01}. Spaces between
several magnetic surfaces $A_y$ = const are shadowed by different
tints. Figure~\ref{fig13}(right) shows the same magnetic surfaces
for the slightly increased value of $I_c$. The magnetic flux
$\Phi$ conservation between the current and the photosphere is
taken into account in the form

\begin{equation}
\Phi = M x_0 \left(\frac{1}{ x_0^2 +
 (z_0 - z_d)^2 } -\frac{1}{ x_0^2 +
 z_d^2 } \right) +  \frac{I}{c} \ln \frac{2 z_0}{r_0}  = const,
\end{equation}
where $r_0 = 0.01$ is the radius of a flux tube with nearly
homogeneous current density, which should be taken into account to
avoid divergency.

When the equilibrium position becomes higher,  some of previously
closed field lines reconnect with open field lines at the null
point. Plasma (possibly previously heated) confined between some
closed  magnetic surfaces is able to propagate along the open 
field lines into the upper corona and to the photosphere. Such 
scenario we expect to happen in the present case of filament
interactions during 2014 April 18 -- 20.

The first event and reconnection on April 18 was initiated by the activation and
partial eruption of SF that bring the inner green field lines 
toward the null point and trigger the reconnection (Figures~\ref{fig12}(b) and~\ref{fig12}(h)). The
observed brightenings (shown by the chartreuse color) near the
foot points of the fan lines on both sides of the filaments is
strong evidence of the null point reconnection (Figure~\ref{fig5}(c)). The accelerated
electrons move after the reconnection towards
the foot points of the fan lines and produce brightenings there.
In the second event on April 19, the activation and partial eruption of NF
bring its field lines to the magnetic null and trigger
reconnection (Figures~\ref{fig12}(c) and~\ref{fig12}(i)). The brightening at the
foot points of the fan lines was also observed in this event. Some
part of the heated NF plasma travels along the outer spine over
the SF towards the western foot point
(Figures~\ref{fig12}(d) and~\ref{fig12}(j)). 
The third event on April 20 is quite 
similar to the second one with a partial eruption of NF again, but there is additional flow of heated
NF plasma along the axis of NF towards the western end
(along the white arrows in Figure~\ref{fig9}(c)). 

Apparent brightening near the null 
point has also been observed just after the partial eruption in both 
the second and third cases, which provides a signature of magnetic reconnection 
(see Figures~\ref{fig8} and~\ref{fig10}). 
In \textit{SDO}/AIA images the filaments and filament channels,
arcades and loops look the same after second event.
However, we believe that there should be some changes in the magnetic configuration
as shown in panel(j) of Figure~\ref{fig12}. The flux rope with helical field lines (black color)
and the outer green spine lines reconnect (Figures~\ref{fig12}(c) and~\ref{fig12}(d)) and create a new domain between a few
new reconnected lines joining the two systems (shown by dotted black lines in
Figures~\ref{fig12}(d) and ~\ref{fig12}(j). We believe that a similar reconnection is occurring 
during the third event (Figure~\ref{fig12}(e) and~\ref{fig12}(k)).
The activation of the filaments can be understood by some photospheric
magnetic changes. Looking at the magnetic field evolution in this region,
we note flux cancellation close to the eastern end of SF, emerging flux close
to the middle and the western end of SF along with the expending motions 
(see the \textit{SDO}/HMI animation attached with Figure~\ref{fig1}).
These motions could create shear and cancelling flux and be the trigger of the
activation of the filaments. The SF remains stable after the first event. From
this we could guess that the twist of SF flux rope became less and the overlying
arcades are more potential. On the other hand the NF being overlaid by less
and more sheared arcades can rise and reach the null point.

\section{Conclusions}
\label{sec6}

In this paper, we discuss the observations of the interactions
between two near-by filament channels of different chirality with parallel axial magnetic fields. We
found a key role of the interaction of partially erupting
filaments and associated flux ropes with the overlying fan-spine 
magnetic structure. On the basis of the analysis of coronal EUV 
images and magnetic field calculations, we come the following main results.

\begin{enumerate}
\item \textit{SDO}/AIA EUV observations show close connection between two
near-by filament channels during three episodes of activation and 
interaction of the filaments with different chiralities on 18, 19, and 20 April 2014, respectively.
\item The observations as well as the PFSS calculations clearly show
the existence of fan-spine  magnetic configuration over the two
filaments.
\item Although the events look like interaction of two
filament channels, the most important interaction occurs between
the flux ropes containing the filaments and the surrounding
coronal magnetic field of special structure. The activations and 
partial eruptions of the filaments are believed to be responsible
for the reconnection between their magnetic field lines with the coronal
field at the magnetic null that lies above them.
\end{enumerate}

The most interesting aspect in these events is the geometry of
the spine line of the coronal fan-spine structure. In projection
on the solar disk, the spine line runs nearly parallel to the SF
axis, close to it. The remote point, where the outer spine line is
anchored in the photosphere, is located near the western end of
SF. It leads to the wrong impression that plasma from one filament
channel easily penetrates to the other filament channel with the
opposite chirality. In fact, the filament plasma penetrates into
the coronal structure and propagates along coronal field lines,
which are located above the SF axis. We strongly believe that the
activations and partial eruptions of both filaments were
responsible for the reconnection at the magnetic null. Plasma of
the filaments then moves along the spine line to the remote
footpoint, which lies near the western end of SF.

Interaction of two nearby filament channels with different filament chiralities within large-scale
coronal fan-spine magnetic structure have not been observed
before, although fan-spine configurations have been discussed in
case of solar eruption in "pseudostreamers" \citep{Wang07,Torok11b,Rachmeler14,Yang15}, 
jets \citep{Pariat09,Filippov09,Filippov15} and flares \citep{Wang14,Joshi15}.

Observations of filament interactions and merging are crucial for
better understanding of reconnection between large-scale coronal
flux ropes. It also provide the information of interaction the flux 
ropes with the overlying magnetic configurations. 
These magnetic structures  are responsible for
different types of eruptions. High-resolution observations provide
important inputs for the MHD modeling of flux-rope interactions
and reconnections, which are needed to understand the physics of
flux-rope dynamics more clearly.


\acknowledgments The authors thank the referee for his/her valuable comments/suggestions. 
We thank SDO/AIA, SDO/HMI, BBSO and GONG/NSO
teams for providing their data for the
present study. This work is supported by the BK21 plus program
through the National Research Foundation (NRF) funded by the
Ministry of Education of Korea. NCJ thank School of Space
Research, Kyung Hee University for providing Postdoctoral grant. We are thankful to Dr. Pascal D{\'e}moulin for his valuable suggestions.





\clearpage


\clearpage


\clearpage


\clearpage


\clearpage


\clearpage


\clearpage


\clearpage


\clearpage


\clearpage


\clearpage


\clearpage


\clearpage


\end {document}